# Hydration Free Energies of Linear Alkanes: Systematic Deviations in Common Water Models and Their Correction

Yalda Ramezani and Sumit Sharma*

**ABSTRACT:** Common force fields overestimate the hydration free energies of hydrophobic solutes, leading to an exaggerated hydrophobic effect. We compute the hydration free energies of linear alkanes from methane to eicosane ($C_{20}H_{42}$) using free energy perturbation with various three-site (SPC/E, OPC3) and four-site (TIP4P/2005, OPC) water models in combination with the TraPPE-UA alkane force field. All water models overestimate the hydration free energies, although the four-site models perform better than the three-site ones. Using alkane cavity free energies, we reparameterize the alkane−water Lennard−Jones well-depth to bring the simulation results in agreement with experimental and group-contribution estimates at 300 K. The reparameterized models significantly improve agreement with experiments across temperatures (290−350 K). We also show that the General Amber force field (GAFF) with TIP4P/2005 water provides closer agreement with experimental hydration free energies than the original TraPPE-UA/TIP4P/2005 combination. Finally, we show that applying a shifted Lennard−Jones potential introduces systematic deviations in the hydration free energies.

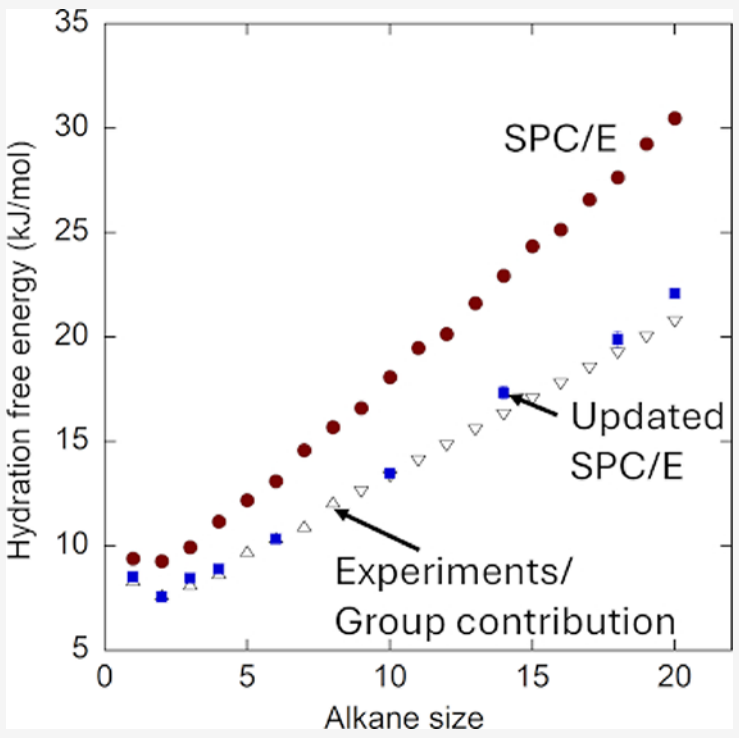

## 1. INTRODUCTION

The hydrophobic effect is a fundamental driving force in many chemical and biological processes, including the self-assembly of surfactants and lipids into micelles and membranes, the folding and stability of proteins, the binding of ligands into hydrophobic pockets of proteins, and the coalescence of oil droplets in water.[1−3] Recent molecular simulation studies have noted that the commonly used force fields tend to exaggerate the hydrophobic effect, leading to underestimated critical micelle concentrations of surfactants and overestimated oil−water interfacial adsorption free energies.[4−6] These systematic deviations highlight the need for improved parameterizations of existing force fields.

Alkanes in water serve as a model system to understand the behavior of nonpolar species in aqueous environments. A key property of interest is their solubility, which can be quantified by the hydration free energy or the change in free energy when an alkane transfers from the vapor to the aqueous phase. The hydration free energy can be estimated using molecular simulations, but the accuracy depends on both the choice of molecular force fields and the numerical implementation details. Molecular force fields are often parametrized on pure species thermodynamic properties. The interspecies force field parameters are then obtained by heuristic mixing rules. The Lorentz−Berthelot mixing rule is the most popular among them. The numerical implementation details refer to the chosen spatial potential cutoff and how the interactions are accounted for beyond the cutoff.[7] In this work, we show that various water models overestimate the hydration free energies of linear alkanes (modeled via TraPPE-UA) when standard Lorentz−Berthelot mixing rules are used. Using the cavity free energies of alkanes, we provide a one-step mechanism to adjust the alkane−water Lennard−Jones well depth parameter to reproduce the experimental values. The reparameterized force field reproduces the experimental hydration free energies across a range of temperatures.

Nonpolar species, like alkanes, do not strongly interact with water due to their inability to form hydrogen bonds and their charge neutrality. Such species disrupt the molecular arrangement of water in their vicinity, making their dissolution in water unfavorable. Thus, nonpolar species tend to aggregate in aqueous environments, which is referred to as the hydrophobic effect.[1] The hydrophobic effect is length-scale dependent.[8,9] At small length-scales, where the size of the solute is less than 1 nm, the hydrophobic effect is entropically driven,[8,10,11] and the hydration free energy is proportional to the solute volume.[9] Hydration free energies of small solutes and their aggregation in water are well-described by the observation that water density fluctuations are Gaussian in small observation volumes.[12] For larger solutes, the hydrophobic effect is dominated by the enthalpic loss of hydrogen bonds of water in their vicinity, and the hydration free energy scales

**Received:** March 7, 2026
**Revised:** April 9, 2026
**Accepted:** April 13, 2026
**Published:** April 16, 2026

proportionally to the exposed surface area of the solute.[9,13] Near large hydrophobic solutes, the low-density solvent fluctuations are enhanced relative to Gaussian statistics.[14,15] Thus, water near large hydrophobic solutes sits at the edge of a dewetting transition.[16,17] As a result, liquid water, when confined between large nonpolar solutes, becomes metastable with respect to its vapor below a critical confinement gap.[18−21] This phenomenon is understood to dictate the hydrophobic self-assembly of large solutes.

The Gibbs free energy of solvation is given by $\Delta G_{hyd} = \Delta H_{hyd} - T\Delta S_{hyd}$, where $\Delta H_{hyd}$ is the change in enthalpy when the solute enters the aqueous phase, and $\Delta S_{hyd}$ is the associated change in entropy. $\Delta H_{hyd}$ comprises solute−solvent direct interactions, $\Delta E_{UV}$, and changes in the solvent−solvent interactions, $\Delta H_{VV}$ ($=\Delta E_{VV} + P\Delta V$). The $\Delta H_{VV}$ exactly cancels out the solvent's entropic contribution, $T\Delta S_{VV}$.[22] Therefore, the $\Delta H_{hyd}$ is determined entirely by the solute−solvent contribution, that is, $\Delta G_{hyd} = \Delta E_{UV} - T\Delta S_{UV}$.[22] Previous works have shown that the positive (unfavorable) $\Delta G_{hyd}$ of alkanes arises from a near-perfect cancellation of the attractive (favorable) solute−solvent interactions, $\Delta E_{UV}$, and the unfavorable solute−solvent entropic term, $T\Delta S_{UV}$.[22] The $\Delta G_{hyd}$ can also be broken down into the free energy of creating a solute-sized cavity, $\Delta G_{cavity}$, and the energetic contribution from the attractive solute−solvent interactions, $\Delta H_{att}$.[23]

Several experiments have reported the solubility of linear alkanes in water, but reliable estimates are available only up to decane because of the extremely low solubilities of longer alkanes.[24−32] Cabani et al.[33] developed a group contribution method based on the experimentally known solubilities of small alkanes to predict the solubilities of linear- and cyclo-alkanes. This group contribution method was updated for aliphatic and monoaromatic hydrocarbons, monohydric alcohols, and aliphatic, noncyclic ketones.[34,35] In the group contribution method, the thermodynamic property of interest is correlated with the number and types of molecular fragments that make up a molecule, allowing the estimation of the average contribution of each molecular fragment toward that thermodynamic property.

Aqueous solubility of alkanes has been the subject of several molecular simulation studies. Ferguson et al.[36] employed the Transferable Potentials for Phase Equilibria (TraPPE)- United Atom (UA) force field[37] for alkanes and the Simple Point Charge Enhanced (SPC/E) water model.[38] Using the incremental Widom insertion technique[39] and a potential cutoff of 9.875 Å for both Lennard−Jones and electrostatic interactions, they determined the aqueous solubility of *n*-alkanes up to docosane ($C_{22}$). While Ferguson et al.'s[36] calculations matched experimental values, other studies that employed the same molecular potentials reported deviations from the experiments.[40,41]

Ashbaugh et al.[41] compared ten different water models for their ability to reproduce the experimental temperature-dependence of liquid water density and methane hydration and concluded that the TIP4P/2005 water model performs best. Ashbaugh et al.[42] calculated the hydration free energy of linear and branched alkanes using the TraPPE-UA + TIP4P/2005 potentials for alkanes and water, respectively, via thermodynamic integration, and concluded that the hydration free energies are overestimated compared to the experimental values when the standard Lorentz−Berthelot mixing rules are used. They adjusted the Lennard−Jones well-depth parameter between the alkane united atoms and water oxygen to match the experimental data, while the Lennard−Jones distance parameter $\sigma$ was set to fix the thermal radius of all sites at 300 K. The new model, called HH-Alkane, reduced the deviation of the hydration free energy from experiments to 0.25 kJ/mol on average. However, their simulations were performed for alkanes of size up to butane and neopentane.

Chen and Siepmann[43] studied the hydration of small alkanes via Gibbs ensemble Monte Carlo simulations using the Optimized Potentials for Liquid Simulations (OPLS) force field[44] for alkanes and the TIP4P model of water. They reported that the hydration free energy of alkanes from the simulations was consistently higher than experimental values. Similar conclusions were drawn in another study by Siepmann and co-workers,[40] where they employed the TraPPE force field for *n*-alkanes along with different water models: TIP4P, TIP4P/2005, and SPC/E. Interestingly, though, they found that the hydration free energies are closer to experimental values for the SPC/E water model compared to those of the TIP4P/2005 water model. Singh and Sharma[45] reported large deviations in the hydration free energies of *n*-alkanes from experimental values in their simulations. They employed the General Amber Force field (GAFF)[46] for alkanes and SPC/E water with a spherical cutoff of 10 Å, and the potentials were shifted by their value at the cutoff distance.

In this work, we calculate the hydration free energies of linear alkanes from methane up to eicosane ($C_{20}H_{42}$) using the TraPPE-UA force field for alkanes combined with various three- and four-point water models. All tested water models systematically overestimate the hydration free energies. The "Optimal 3-Charge, 4-Point rigid" (OPC) water model[47] and its 3-site counterpart OPC3,[48] were designed by optimizing charge distributions to capture bulk electrostatics, yielding hydration free energies similar to those from TIP4P/2005 and SPC/E, respectively. By utilizing the free energy of cavity formation, we adjust the alkane−water Lennard−Jones well-depth parameter ($\epsilon$) for each model to match experimental values at 300 K. Remarkably, in all cases, the well-depth parameter needs to be increased by about 5% relative to the Lorentz−Berthelot mixing rule. The reparametrized models show good agreement with the experimental hydration free energies at different temperatures. We do not adjust the Lennard−Jones distance parameter $\sigma$, as the cavity free energy is a strong function of $\sigma$. We show that the HH-alkane model with TIP4P/2005 water reproduces experimental/group contribution hydration free energies up to eicosane. The HH-alkane model was not tested before for linear alkanes larger than butane.[42] We also find that the hydration free energies obtained using the GAFF with TIP4P/2005 and SPC/E water models result in smaller deviations from experiment compared to the TraPPE-UA model. Finally, we demonstrate that shifting the Lennard−Jones potential by its value at the spherical cutoff increases deviations from the experimental results. More broadly, these results highlight that molecular force fields parameterized on pure-component thermodynamics often perform poorly for mixtures when heuristic mixing rules like Lorentz−Berthelot are used. For nonpolar species like alkanes, accurate hydration free energies can be obtained by directly calibrating the interaction parameters using cavity-formation free energies.

## 2. SIMULATION SYSTEM AND METHODS

A widely used force field for alkanes is the Transferable Potentials for Phase Equilibria (TraPPE)- United Atom (UA) model.[37] In this

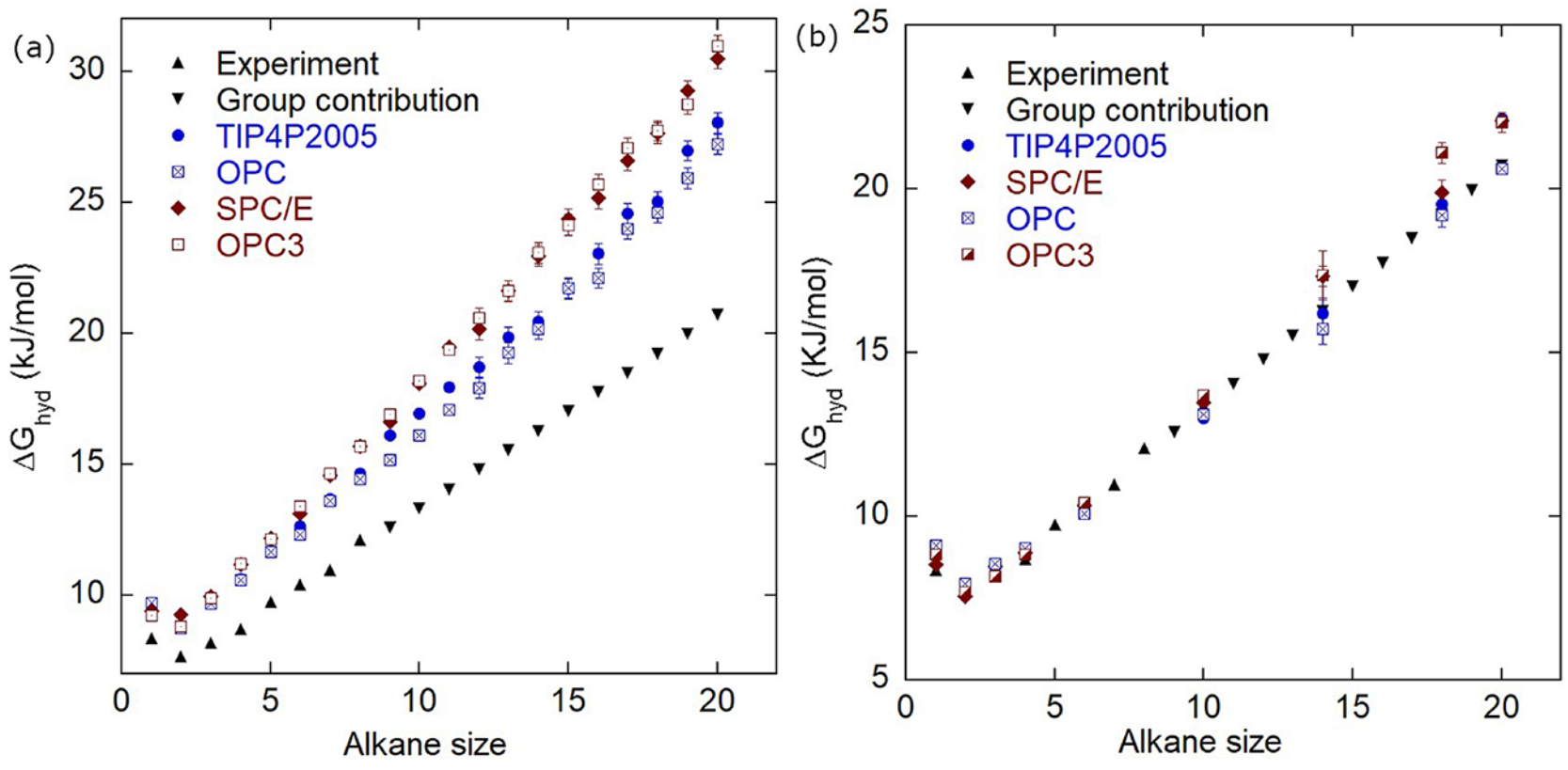

**Figure 1.** (a) Hydration free energies, $\Delta G_{hyd}$, of alkanes for TraPPE-UA and different water models. All models show positive deviation from experiments/group contribution. The 3-point water models show larger deviation compared to the 4-point water models. (b) $\Delta G_{hyd}$ of alkanes using the TraPPE-UA model but with the alkane−water well-depth parameter updated via the method described in the text. An excellent match with experimental/group-contribution values is found, except for some deviations observed with the OPC3 model for large alkanes. The uncertainty in experimental $\Delta G_{hyd}$ is 0.84 kJ/mol for alkanes up to pentane and 2.51 kJ/mol for alkanes up to decane.[58]

model, each alkyl group is represented as a single united atom bead with distinct sites defined for $CH_4$, $-CH_3$, and $-CH_2-$. Beyond alkanes, TraPPE-UA has also been parameterized for alcohols, thiols, ethers, sulfides, aldehydes, ketones, nitriles, cycloalkanes, and aromatics. In this work, we evaluate the performance of TraPPE-UA with the GAFF[46] and the HH-alkane model.[42] We employ four commonly used models: single point charged enhanced (SPC/E), optimal point charge 3 (OPC3), TIP4P/2005, and OPC. SPC/E and OPC3 are three-site models, while TIP4P/2005 and OPC are four-site models, in which the negative charge is shifted off the oxygen atom onto the H−O−H angle bisector.[47−49]

Hydration free energies of alkanes are calculated using the free energy perturbation (FEP) method.[4,50] The alkane−water interactions are described by a soft-core Lennard−Jones potential (eq 1),[51] with $\alpha = 0.5$ and $n = 2$.

$$V(\lambda, r) = \lambda^n 4\epsilon\left[\left(\alpha(1-\lambda)^2 + \left(\frac{r}{\sigma}\right)^6\right)^{-2} - \left(\alpha(1-\lambda)^2 + \left(\frac{r}{\sigma}\right)^6\right)^{-1}\right] \quad (1)$$

The coupling parameter $\lambda$ is varied from 0 to 1 in 40 windows with $\Delta\lambda = 0.025$. Each window is simulated for 6 ns, and ensemble averages are computed from the last 4.5 ns. Accuracy of FEP is validated by tracing the reverse path ($\lambda$ = 1 to 0) using the same protocol. Reported uncertainties are the standard deviation over three independent simulations.

Hydration free energy is calculated in the isothermal−isobaric ensemble using eq 2.[52,53]

$$\Delta G = -k_B T \sum_{i=0}^{n-1} \ln\left(\frac{\left\langle V \exp\left(-\frac{U(\lambda_{i+1}) - U(\lambda_i)}{k_B T}\right)\right\rangle_{\lambda_i}}{\langle V\rangle_{\lambda_i}}\right) \quad (2)$$

The simulation system is composed of one alkane molecule in bulk water. The nominal size of the cubic simulation system is selected to ensure that the periodic images of the alkane do not interact with each other (Table S1, Supporting Information). Periodic boundary conditions are applied in all directions. Coulombic interactions between water molecules are calculated using the Particle−Particle Particle-Mesh Ewald (PPPM) summation with an accuracy of $10^{-4}$.[54] Lennard−Jones and the real-space part of the Coulombic interactions between water molecules is truncated at a spherical cutoff of 10 Å. The Lennard−Jones interactions between water-oxygen and alkane atoms have a spherical cutoff of 14 Å, as recommended in the TraPPE-UA force field.[55] Unless noted otherwise, the potential functions are not shifted by their value at the spherical cutoff, and long-range/tail corrections to the potential and pressure are applied for the Lennard−Jones interactions. The simulations are performed in the isothermal−isobaric ensemble at a pressure $P$ = 1 bar. Temperature and pressure in the system are maintained using the Nose−Hoover thermostat and barostat, respectively. All simulations are conducted using the Large-scale Atomic/Molecular Massively Parallel Simulator (LAMMPS).[56] Initial configurations are generated using the PACKMOL package.[57]

## 3. RESULTS AND DISCUSSION

Figure 1a presents the hydration free energies, $\Delta G_{hyd}$, of alkanes from methane to eicosane, calculated using the TraPPE-UA alkane potential with four different water models (SPC/E, OPC3, TIP4P/2005, and OPC) at temperature $T$ = 300 K. The results are compared with experimental $\Delta G_{hyd}$ values up to octane and with estimates from the group contribution method[33] for longer alkanes. Experimental $\Delta G_{hyd}$ is tabulated in Cabani et al.[33] and the Freesolv database.[58] The uncertainty in experimental $\Delta G_{hyd}$ is 0.84 kJ/mol for alkanes up to pentane and 2.51 kJ/mol for alkanes up to decane.[58] For all water models, the TraPPE-UA force field overpredicts $\Delta G_{hyd}$. The deviations are larger for the 3-point water models (SPC/E and OPC3) than for the 4-point water models (TIP4P/2005, OPC), and the deviations increase with alkane length. The two 3-point water models yield similar results, as do the two 4-point water models. Our results are consistent with the observations by Kanduč et al.[4] and others,[5,6] who report that the currently available atomistic force fields systematically yield stronger adsorption affinities for oil−water interfaces and under-predict the critical micelle concentration. Our $\Delta G_{hyd}$ values match well with the estimates in previous studies, as shown in Table S2 (Supporting Information).

To reconcile the hydration free energies obtained in simulations with the experiments, we reparameterize the alkane−water Lennard−Jones well-depth parameter, $\epsilon$. Our methodology is as follows. The hydration free energy can be expressed as $\Delta G_{hyd} = \Delta G_{cavity} + \Delta H_{att}$, where $\Delta H_{att} \approx \Delta U_{att}$ corresponds to the alkane−water Lennard−Jones interaction energy. The approximation $\Delta H_{att} \approx \Delta U_{att}$ is justified because the volume change associated with transferring an alkane into water is mainly accounted for in the $\Delta G_{cavity}$ term.[59] Values of $\Delta G_{cavity}$ are taken from previous studies.[45,59] Our calculation of $\Delta G_{cavity}$ at 350 K shows that $\Delta G_{cavity}$ does not change in this temperature range (Figure S1, Supporting Information). The weak temperature dependence of the cavity free energy reflects the limited variation of the water density, pair correlation function, and associated density fluctuations over the temperature range considered. The experimental (or group contribution) estimate of the $\Delta H_{att}$ can then be obtained as $\Delta H_{att}^{exp} = \Delta G_{hyd}^{exp} - \Delta G_{cavity}$. This way, $\Delta H_{att}$ can be computed for each alkane + water model combination. For small perturbations to $\epsilon$, the local solvent structure around the alkane is assumed to be unchanged. Therefore, the updated $\epsilon$ is found as $\epsilon^{updated} = \frac{\Delta H_{att}^{exp}}{\Delta H_{att}} \times \epsilon$. This procedure is demonstrated for the SPC/E water model in Table 1. Interestingly, for all models, $\epsilon$ needs to be increased by approximately 5% relative to its Lorentz–Berthelot value. Applying it to all water models yields the updated $\epsilon$ values summarized in Table 2. The $\sigma$ parameters of the HH-alkane model are also reported for comparison.[42] In the HH-alkane model, the Lennard–Jones distance parameter $\sigma$ is set so that the thermal radius of each site is fixed at 300 K. Figure 1b shows that $\Delta G_{hyd}$ obtained using the updated $\epsilon$ for all the water models has an excellent match with experiments/group contribution values at $T = 300$ K, except for OPC3, which shows some deviation for large alkanes. The change in the free energy as a function of the FEP parameter $\lambda$ is shown in Figure S2 (Supporting Information).

**Table 1. Comparison of Experimental Hydration Data for TraPPE-UA + SPC/E and the Updated SPC/E Models[a]**

| alkane | $\Delta G_{hyd}^{exp}$ | $\Delta G_{hyd}^{SPC/E}$ | $\Delta G_{cavity}$ | $\Delta H_{att}^{exp}$ | $\Delta H_{att}^{SPC/E}$ | $\Delta H_{att}^{SPC/E,updated}$ | $\Delta G_{hyd}^{SPC/E,pred}$ | $\Delta G_{hyd}^{SPC/E,sim}$ |
|---|---|---|---|---|---|---|---|---|
| 1 | 8.37 | 9.39 | 24.52 | −16.15 | −15.12 | −15.96 | 8.56 | 8.53 |
| 2 | 7.66 | 9.28 | 34.87 | −27.21 | −25.59 | −27.01 | 7.86 | 7.57 |
| 3 | 8.18 | 9.94 | 43.85 | −35.67 | −33.91 | −35.78 | 8.07 | 8.47 |
| 4 | 8.70 | 11.16 | 52.55 | −43.85 | −41.39 | −43.68 | 8.87 | 8.89 |
| 5 | 9.76 | 12.19 | 61.86 | −52.10 | −49.67 | −52.42 | 9.44 | |
| 6 | 10.4 | 13.11 | 70.34 | −59.94 | −57.23 | −60.40 | 9.94 | 10.34 |
| 8 | 12.1 | 15.69 | 87.72 | −75.62 | −72.03 | −76.01 | 11.71 | |
| 10 | 13.32 | 18.08 | 105.75 | −92.43 | −87.68 | −92.53 | 13.23 | 13.47 |
| 12 | 14.80 | 20.15 | 123.36 | −108.56 | −103.21 | −108.92 | 14.44 | |
| 14 | 16.28 | 22.95 | 140.80 | −124.52 | −117.85 | −124.37 | 16.43 | 17.34 |
| 16 | 17.76 | 25.15 | 157.81 | −140.05 | −132.66 | −140.00 | 17.81 | |
| 18 | 19.24 | 27.64 | 174.72 | −155.48 | −147.08 | −155.21 | 19.51 | 19.90 |

[a]The units are in kJ/mol. $\Delta G_{hyd}^{SPC/E,pred}$ are the predicted $\Delta G_{hyd}$ values after updating the $\epsilon$. $\Delta G_{hyd}^{SPC/E,sim}$ is the value obtained in the simulations.

**Table 2. Original and Updated Lennard–Jones Well-Depth Parameter ($\epsilon$) for $CH_4$:O, $-CH_3$:O, and $-CH_2$:O for Different Water Models in kJ/mol[a]**

| | $CH_4$:O | | $CH_3$:O | | $CH_2$:O | |
|---|---|---|---|---|---|---|
| | $\epsilon$ | $\epsilon^{updated}$ | $\epsilon$ | $\epsilon^{updated}$ | $\epsilon$ | $\epsilon^{updated}$ |
| SPC/E | 0.8942 | 0.9436 | 0.7276 | 0.7679 | 0.4985 | 0.5261 |
| OPC3 | 0.9172 | 0.9668 | 0.7464 | 0.7867 | 0.5114 | 0.5390 |
| OPC | 1.0467 | 1.0841 | 0.8517 | 0.8822 | 0.5835 | 0.6044 |
| TIP4P2005 | 0.9765 | 1.0169 | 0.7946 | 0.8274 | 0.5444 | 0.5669 |
| HH-alkane | 0.9765 | 1.0226 | 0.7946 | 0.8224 | 0.5444 | 0.5652 |

[a]The Lennard–Jones distance parameter ($\sigma$) was not modified. For comparison, the $\epsilon$ values of the HH-alkane model are also reported.

Next, we assess the accuracy of the parameterized models in reproducing hydration free energies across temperatures. Figure 2 presents $\Delta G_{hyd}$ (in units of $kT$) over the temperature range 290–350 K for the four reparameterized alkane–water models. The reparameterized TIP4P/2005, SPC/E, and OPC3 water models reproduce the experimental data[60] (shown with a black solid line) with overall root mean squared deviations (RMSDs) of 0.07, 0.10, and 0.06 $kT$, respectively. The OPC water model exhibits positive deviations from the experiment, with an overall RMSD of 0.19 $kT$. These deviations can be rationalized by noting that, during parametrization of the OPC model, positive deviations were observed for small alkanes (methane through butane), whereas the predictions for larger alkanes agreed well with experimental/group contribution values (Figure 1). The experimental data itself have an uncertainty of 0.33 $kT$.[58] For comparison, results from the original TraPPE-UA/TIP4P/2005 combination (blue dashed line) are also shown in Figure 2. The original model yields an RMSD of 0.29 $kT$ for methane, and the deviation increases monotonically with the alkane size.

Ashbaugh et al.[42] reported that the TraPPE-UA force field combined with TIP4P/2005 water overestimates the hydration free energies of small alkanes. To address this, they proposed the HH-alkane model. In the HH-alkane model, Ashbaugh et al.[42] set the Lennard–Jones distance parameter $\sigma$ such that the thermal radius of each site is fixed at 300 K and then fit the well-depth parameter $\sigma$ to match $\Delta G_{hyd}$ to the experimental data at different temperatures. They focused on linear alkanes up to butane, isobutane, and neopentane. Here, we extend their analysis by calculating the $\Delta G_{hyd}$ of linear alkanes up to eicosane using the HH-alkane + TIP4P/2005 combination. As shown in Figure 3, the resulting $\Delta G_{hyd}$ values are in excellent agreement with both the experimental/group-contribution data and our reparameterized TraPPE + TIP4P/2005 force field combination at 300 K. Ashbaugh et al.[42] report an overall RMSD of 0.10 $kT$ at different temperatures.

Our approach differs from Ashbaugh et al.[42] in many ways. First, we employ cavity free energies of alkanes to adjust the alkane–water interaction energy at 300 K and then compare $\Delta G_{hyd}$ with experiments at various temperatures. We show that the cavity free energies of alkanes do not change for the highest temperatures studied (350 K), justifying our approach (Figure S1, Supporting Information). Second, we reparameterize for linear alkanes from methane up to eicosane, while Ashbaugh et al. restricted their study to small alkanes. Third, we do not adjust the Lennard–Jones distance parameter $\sigma$ because cavity free energies have a strong dependence on $\sigma$. Ashbaugh et al.[42] estimate excess enthalpy ($\Delta H^{ex}$), excess entropy ($\Delta S^{ex}$) and excess heat capacity ($C^{ex}$) of hydration of alkanes by fitting the following relation:

$$\Delta G_{hyd}(T) = \Delta G_{hyd}(T_0) + (C^{ex} - S^{ex}(T_0))(T - T_0) - C^{ex}T \times \ln(T/T_0) \quad (3)$$

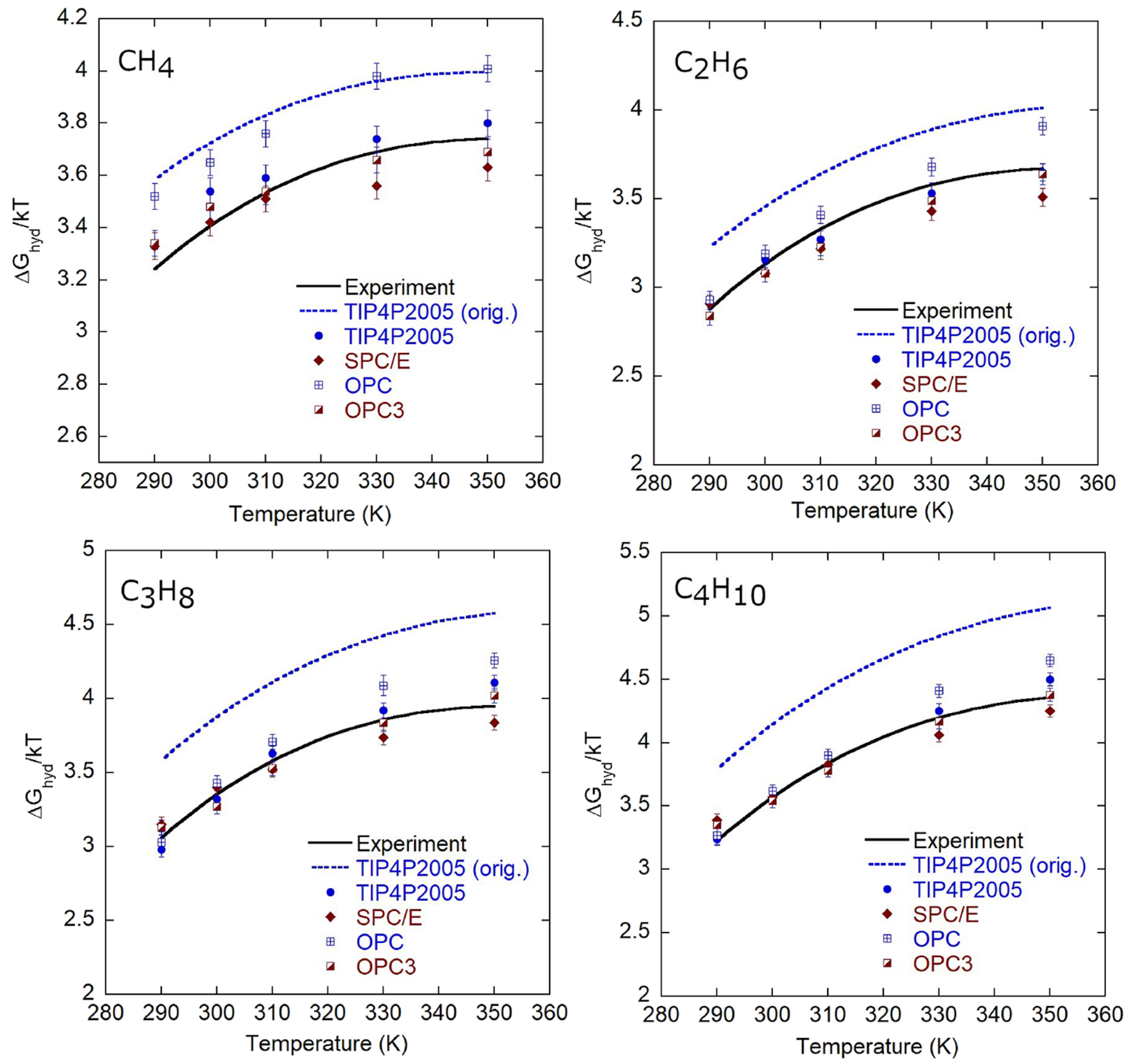

**Figure 2.** Hydration free energies, $\Delta G_{hyd}$, of alkanes from methane to butane over the temperature range 290–350 K. Values are reported in units of $kT$. The black solid line represents experimental data.[60] The blue dashed line corresponds to the original TraPPE-UA/TIP4P/2005 combination. Symbols denote results from the reparameterized models. The reparameterized TIP4P/2005, SPC/E, and OPC3 models yield overall RMSDs of 0.07, 0.10, and 0.06 $kT$, respectively. The OPC model exhibits a positive deviation with an overall RMSD of 0.19 $kT$. The original TraPPE-UA/TIP4P/2005 combination has an RMSD of 0.29 $kT$ for methane, with deviations increasing monotonically with alkane size.

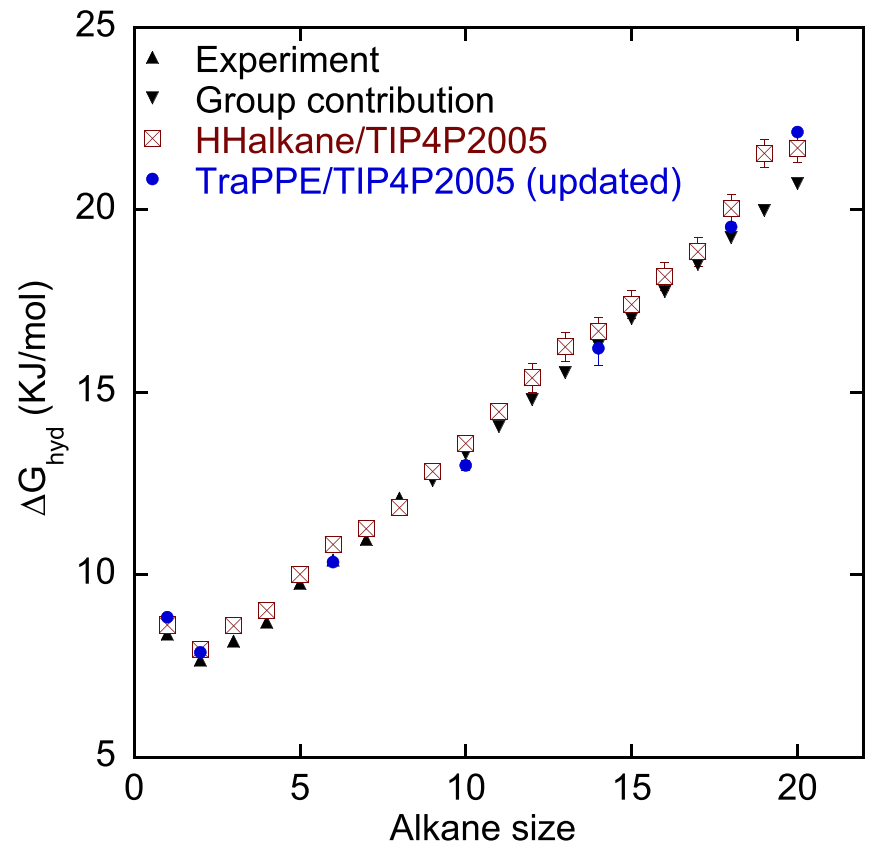

**Figure 3.** Hydration free energies of alkanes calculated using the HH-alkane model at 300 K. Both the HH-alkane and our reparameterized TraPPE/TIP4P/2005 models match the experimental/group contribution values well up to $C_{20}$.

We did not attempt to fit eq 3 to the simulation data because the equation is highly nonlinear in $T$, and the limited temperature range does not allow for reliable fitting.

We also calculated $\Delta G_{hyd}$ using the GAFF with the SPC/E and TIP4P/2005 water models at 300 K. Figure 4 compares the $\Delta G_{hyd}$ obtained from GAFF with those from TraPPE-UA. The alkane–water interactions are determined using the Lorentz–Berthelot mixing rules in both cases. GAFF yields smaller deviations from the experimental and group contribution values compared to TraPPE-UA. Specifically, the GAFF + SPC/E combination systematically overestimates $\Delta G_{hyd}$ by roughly 1–1.6 kJ/mol. GAFF+TIP4P/2005 overestimates $\Delta G_{hyd}$ for small alkanes but underestimates it for larger alkanes. Our results align with those of Luz et al.,[5] who report that GAFF + SPC/E shows too strong an affinity of polyethoxylated alkyl ethers ($C_8EO_m$) to a heptane-water interface.

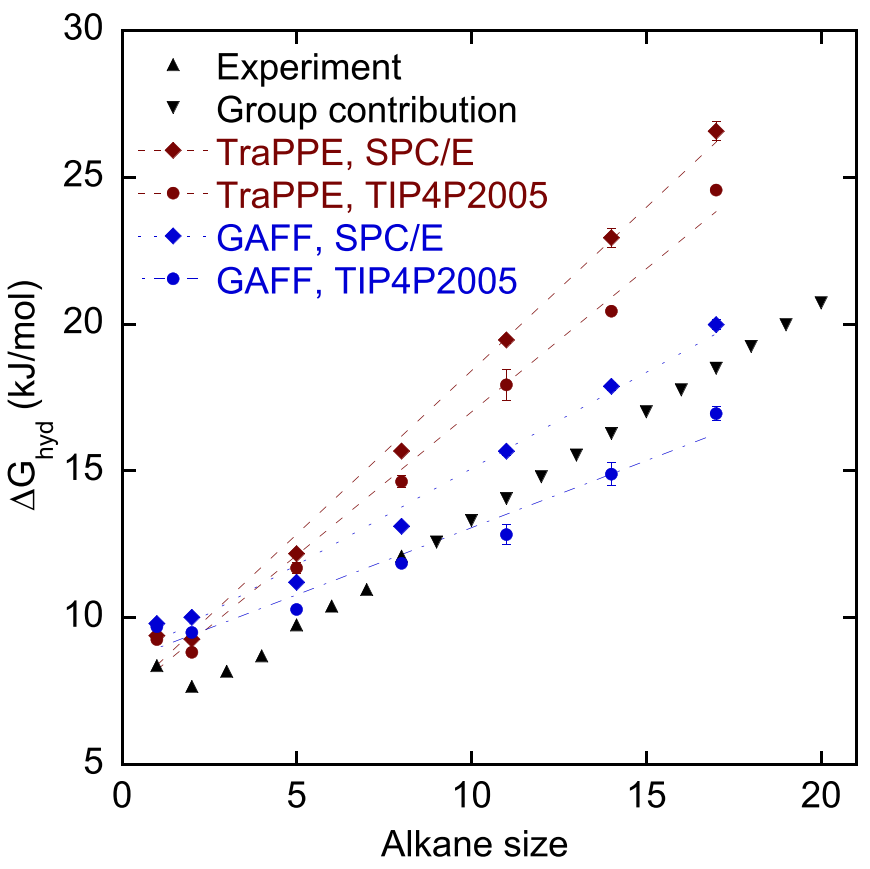

**Figure 4.** Hydration free energies of alkanes calculated using the General Amber Force field (GAFF) all-atom model and two different water models, SPC/E and TIP4P/2005, at 300 K. The results are compared to the TraPPE-UA model and experimental and group contribution values. Overall, results using GAFF are closer to the experimental/group contribution values compared to those of TraPPE-UA. GAFF + SPC/E overestimates $\Delta G_{hyd}$ values. GAFF + TIP4P/2005 overestimates $\Delta G_{hyd}$ for small alkanes ($\leq C_5$) but underestimates $\Delta G_{hyd}$ for large alkanes. Lines are guides to the eyes.

Finally, we discuss the impact of the potential shift on the $\Delta G_{hyd}$ calculations. Often, molecular interaction potentials are shifted by their value at the spherical cutoff so that the potential becomes zero at the cutoff, $r_c$. Shifted potential functions are given by

$$u_{shift}(r_{ij}) = \begin{cases} u(r_{ij}) - u(r_c), & r_{ij} \leq r_c \\ 0, & r_{ij} > r_c \end{cases} \tag{4}$$

For the "unshifted" potentials, the contribution of the potential (which remains nonzero) beyond $r_c$, called the long-range or tail correction, is included for energy and pressure calculations. It should be noted that in molecular dynamics, only the forces matter. So whether a potential function is shifted or not and whether a tail correction is applied does not influence the dynamics and final trajectories, except through pressure calculations, which could affect the NPT ensemble.

For molecules belonging to different species (for example, an alkane molecule in water), the expression for the ensemble-averaged interaction energy is given by

$$\langle E \rangle = N_{alkane}\rho_{water} \int_0^\infty 2\pi r^2 g(r) u(r) dr \tag{5}$$

where $N_{alkane}$ is the number of united atoms in the alkane molecule, $\rho_{water}$ is the number density of water, and $g(r)$ is the alkane−water radial distribution function. Change in the energy due to the shifting of the potential by $u(r_c)$ is given by

$$\langle \Delta E \rangle = N_{alkane}\rho_{water} u(r_c) \int_0^{r_c} 2\pi r^2 g(r) dr \tag{6}$$

eq 6 can be approximated as,

$$\langle \Delta E \rangle = N_{alkane}\rho_{water} u(r_c) \left[ \frac{2}{3}\pi((r_c - \sigma)^3 - (\tilde{N}_{alkane} - 1)\sigma^3) \right] \tag{7}$$

In eq 7, the first term in the brackets assumes $g(r) = 1$ for $r > \sigma$ and 0 for $r < \sigma$. The second term accounts for the excluded volume of other alkane united atoms. $\tilde{N}_{alkane}$ refers to the number of alkane united atoms that are within the cutoff distance, $r_c$, of one united atom. For small alkanes, $\tilde{N}_{alkane} = N_{alkane}$. For the spherical cutoff, $r_c = 1.4$ nm, $\tilde{N}_{alkane} = 11$.

The tail contribution to the energy in the case of an "unshifted" potential is also calculated through eq 5 with the assumption that $g(r) = 1$ for $r > r_c$. The tail contribution of the Lennard−Jones interactions between alkane and water is given by

$$\langle E_{LR} \rangle = N_{alkane}\rho_{water} 8\pi\epsilon\sigma^3 \left[ \frac{\sigma^9}{9r_c^9} - \frac{\sigma^3}{3r_c^3} \right] \tag{8}$$

Figure 5 compares the $\Delta G_{hyd}$ obtained for the unshifted HH-alkane potential (labeled as HHalkane), the shifted HH-alkane (labeled as HHalkane (shifted)), the shifted HH-alkane with the effect of potential shift corrected usingeq 7 and the tail corrections added (labeled as HHalkane (shift adj., tail), and the shifted HH-alkane with only the tail corrections added (labeled as shifted, only tail). The main observation is that the shifted potential exacerbates the deviation from the experimental/group contribution values. The correction for the shifted potential + the tail correction results in a negative deviation from the $\Delta G_{hyd}$ obtained from the unshifted potential, with the largest deviation of 3.1 kJ/mol observed for $C_{19}$. Adding only the tail correction to the shifted potential matches the results of the unshifted potential, but this is just fortuitous.

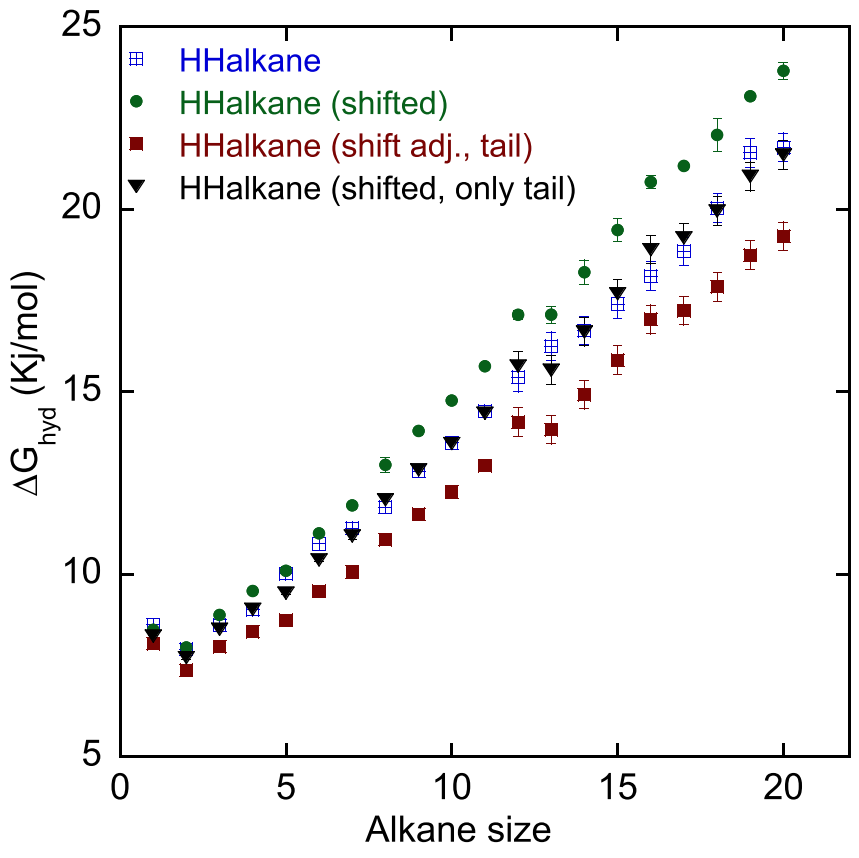

**Figure 5.** Hydration free energies, $\Delta G_{hyd}$, of alkanes for the unshifted HH-alkane (HHalkane) compared to those obtained for the shifted HH-alkane (HHalkane (shifted)); the shifted HH-alkane with the potential shift corrected using eq 7 and tail corrections added (eq 8) (HHalkane (shift adj., tail)); and the shifted potential with only tail corrections added (HHalkane (shifted, only tail)).

Figure 6 shows the carbon−oxygen radial distribution functions $g(r)$ for four different alkane−water systems. The $g(r)$ values of $C_{10}$, $C_{14}$, and $C_{20}$ are shifted vertically by 0.1 for ease of visualization. Beyond a weak first peak and a trough $g(r)$ is close to 1. Thus, the assumption in eq 7 of $g(r) = 1$ for $r > \sigma$ is justified. Table 3 compares the contribution of the potential shift on the $\Delta G_{hyd}$ calculated using eqs 7 and 6 for the case of HH-alkane with TIP4P/2005 water with a spherical cutoff of 1.4 nm. The table shows that the two estimates are close, justifying the approximation in eq 7.

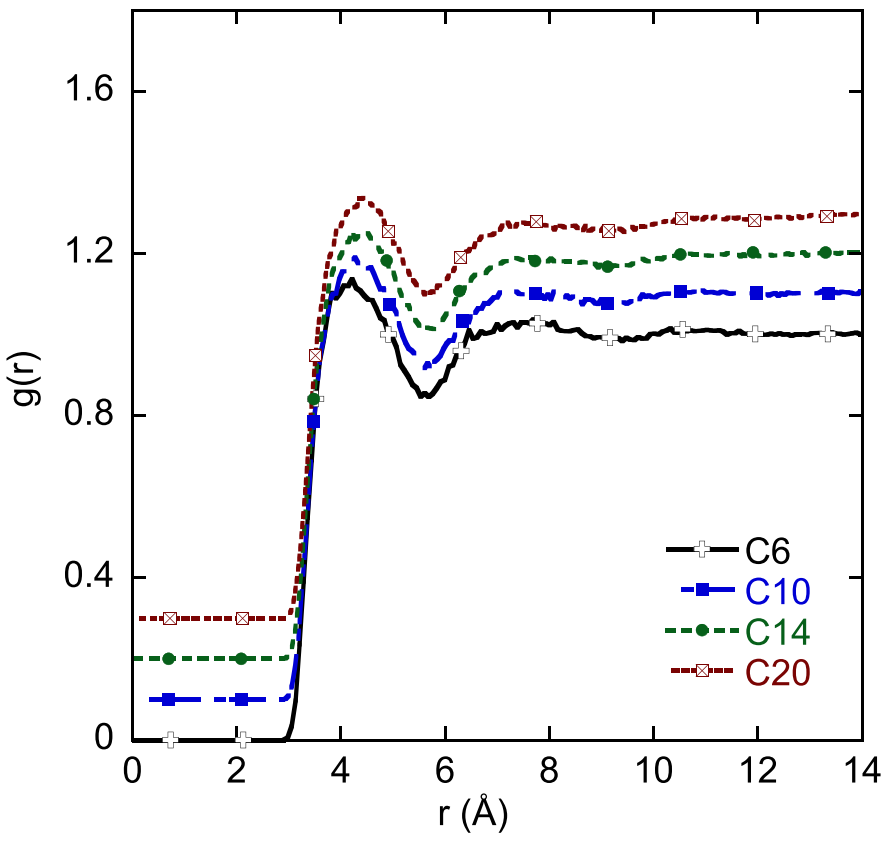

**Figure 6.** Carbon−oxygen radial distribution functions for four different alkanes. The radial distribution functions of $C_{10}$, $C_{14}$, and $C_{20}$ are shifted vertically by 0.1, 0.2, and 0.3, respectively, for ease of visualization.

**Table 3. Contribution of the Potential Shift to the Hydration Free Energies, $\Delta G_{hyd}$, of Different Alkanes When the HH-Alkane Force Field with TIP4P/2005 Water Is Used with a Spherical Cutoff of 1.4 nm**[a]

| alkane | $\Delta E$ (eq 7) (kJ/mol) | $\Delta E$ (eq 6) (kJ/mol) |
|---|---|---|
| 6 | −0.87 | −0.73 |
| 10 | −1.34 | −1.17 |
| 14 | −1.82 | −1.60 |
| 20 | −2.58 | −2.24 |

[a]The second column is the approximation based on eq 7. Values in the third column are calculated from the radial distribution functions using eq 6. The two estimates differ by only 0.33 kJ/mol at most.

## 4. CONCLUSIONS

We show that the hydration free energies of linear alkanes from methane to eicosane ($C_{20}H_{42}$) computed using the TraPPE-UA alkane model and four water models (TIP4P/2005, OPC, SPC/E, and OPC3) are systematically overestimated relative to experimental/group contribution values when Lorentz−Berthelot mixing rules are applied. Using the cavity free energies of alkanes, we adjust the alkane−water Lennard−Jones well-depth parameter ($\epsilon$) to match the hydration free energies to experimental/group-contribution values at 300 K. For each water model, the optimal $\epsilon$ is approximately 5% greater than its Lorentz−Berthelot value. The reparameterized models show good agreement with the experimental hydration free energies at different temperatures. We compare the predictions from the GAFF all-atom alkane model to those of the TraPPE-UA model. The GAFF potential exhibits smaller deviations from experimental/group contribution values. Lastly, we show that applying shifted potential functions increases the deviation of the hydration free energies from the experimental and group contribution values.

## ■ ASSOCIATED CONTENT

### Data Availability Statement

The data that support the findings of this study are openly available in Zenodo at 10.5281/zenodo.15875249.[61]

### Supporting Information

The Supporting Information is available free of charge at https://pubs.acs.org/doi/10.1021/acs.jpcb.5c07832.

Table S1 lists details of the alkane-water simulation systems; Table S2 compares our hydration free energy estimates with those reported in previous studies at 300 K; Table S3 lists $\Delta G_{hyd}$ from experiments, group-contribution, and all the four water models with alkane-water interaction parameters estimated using the Lorentz−Berthelot mixing rules at 300 K; Table S4 lists the $\Delta G_{hyd}$ obtained for the four water models with the updated alkane-water well-depth parameter ($\epsilon$) at 300 K; Table S5 lists $\Delta G_{hyd}$ of alkanes from methane to butane for temperatures ranging from 270−370 K, calculated for different water models with updated alkane-water well-depth parameter; Table S6 lists the $\Delta G_{hyd}$ from GAFF with TIP4P/2005 and SPC/E at 300 K; Figure S1 compares the cavity free energy of alkanes at 300 and 350 K; Figure S2 shows change in the free energy in FEP simulations as a function of the FEP parameter, $\lambda$ (PDF)

## ■ AUTHOR INFORMATION

### Corresponding Author

**Sumit Sharma** − *Department of Chemical and Biomolecular Engineering, Ohio University, Athens, Ohio 45701, United States;* orcid.org/0000-0003-3138-5487;
Email: sharmas@ohio.edu

### Author

**Yalda Ramezani** − *Department of Chemical and Biomolecular Engineering, Ohio University, Athens, Ohio 45701, United States;* orcid.org/0009-0008-7596-8987

Complete contact information is available at:
https://pubs.acs.org/10.1021/acs.jpcb.5c07832

### Author Contributions

S.S. conceptualized the project. Y.R. and S.S. designed the methodology. Experiments were performed and analyzed by Y.R.; Y.R., and S.S. wrote the original manuscript draft, and performed editing and review. S.S. supervised the project and obtained resources.

### Notes

The authors declare no competing financial interest.

## ■ ACKNOWLEDGMENTS

This work is supported by the National Science Foundation (NSF) CAREER grant 2046095. Computational resources for this work were provided by the NSF ACCESS grant DMR190005 and NSF MRI grant 2320493. The authors acknowledge detailed discussions with Drs. Amish Patel, Matej Kanduč, Dor Ben-Amortz, Hank Ashbaugh, Michael R. Shirts, Andrew Ferguson, Ilja Siepmann, and Himanshu Singh.

## ■ REFERENCES